\documentclass[aps,prb,twocolumn,superscriptaddress,showpacs]{revtex4}
\usepackage{graphicx}
\usepackage{amssymb}
\usepackage{bm}

\begin{document}


\title{Disorder suppression and precise conductance quantization
in constrictions of PbTe quantum wells}



\author{G. Grabecki}
\email[]{grabec@ifpan.edu.pl}
\affiliation{Institute of Physics, Polish Academy of Science,
al.~Lotnik\'ow 32/46, 02-668 Warszawa, Poland}
\author{J. Wr\'obel}
\affiliation{Institute of Physics, Polish Academy of Science,
al.~Lotnik\'ow 32/46, 02-668 Warszawa, Poland}
\author{T. Dietl}
\affiliation{Institute of Physics, Polish Academy of Science,
al.~Lotnik\'ow 32/46, 02-668 Warszawa, Poland}
\affiliation{Institute of Theoretical Physics, Warsaw University,
00-681 Warszawa, Poland}
\affiliation{ERATO
Semiconductor Spintronics Project, Japan Science and Technology
Agency, al.~Lotnik\'ow 32/46, 02-668 Warszawa, Poland}
\author{E. Janik}
\affiliation{Institute of Physics, Polish Academy of Science,
al.~Lotnik\'ow 32/46, 02-668 Warszawa, Poland}
\author{M. Aleszkiewicz}
\affiliation{Institute of Physics, Polish Academy of Science,
al.~Lotnik\'ow 32/46, 02-668 Warszawa, Poland}
\author{E. Papis}
\address{Institute of Electron Technology,
al. Lotnik\'ow 32/46, 02-668 Warszawa, Poland}
\author{E. Kami\'nska}
\address{Institute of Electron Technology,
al. Lotnik\'ow 32/46, 02-668 Warszawa, Poland}
\author{A. Piotrowska}
\address{Institute of Electron Technology,
al. Lotnik\'ow 32/46, 02-668 Warszawa, Poland}
\author{G. Springholz}
\address{Institut fu\"r Halbleiter- und Festko\"rperphysik, Johannes
Kepler Universit\"at, A-4040 Linz, Austria}
\author{G. Bauer}
\address{Institut fu\"r Halbleiter- und Festko\"rperphysik, Johannes
Kepler Universit\"at, A-4040 Linz, Austria}

\date{\today}

\begin{abstract}
Conductance quantization was measured in submicron constrictions
of PbTe, patterned into narrow,12 nm wide quantum wells deposited
between Pb$_{0.92}$Eu$_{0.08}$Te barriers. Because the quantum
confinement imposed by the barriers is much stronger than the
lateral one, the one-dimensional electron energy level structure
is very similar to that usually met in constrictions of
AlGaAs/GaAs heterostructures. However, in contrast to any other
system studied so far, we observe precise conductance
quantization in $2e^2/h$ units, {\it despite of significant
amount of charged defects in the vicinity of the constriction}.
We show that such extraordinary results is a consequence of the
paraelectric properties of PbTe, namely, the suppression of
long-range tails of the Coulomb potentials due to the huge
dielectric constant.
\end{abstract}


\maketitle

\section{Introduction}


Present state-of-the art epitaxial growth and processing
techniques enables one to fabricate semiconductor nanostructures
whose dimensions are comparable to the de Broglie wavelength of
band carriers. Typical examples are narrow point contacts whose
conductance becomes quantized in $2e^2/h$ units,\cite{vWee88} as
well as quantum dots revealing discrete energy levels, in close
analogy to atomic levels.\cite{Taru96} Recently, much research
effort has been devoted to quantum nanostructures,  as they could
form the hardware basis of quantum information and communication
technologies.\cite{Loss98, hans05} However, one of the problems
to be solved prior to practical implementation is control over
electrostatic potential on the nanometer-scale.\cite{galp03}
Random potential fluctuations in nanostructures are produced
either by unintentional defects introduced during the processing
or heteroepitaxial growth or by artificially incorporated doping
impurities. The latter cannot be completely avoided as they
provide the free carriers necessary for the device operation.
Modulation doping, in which the doped region is spatially
separated from the free carriers has been utilized as a means to
elevate the problem and accordingly, up to now most of
experimental results have been obtained for modulation doped
AlGaAs/GaAs nanostructures\cite{cron02, grah03}. However, even
this almost perfect and developed system is not free of random
potential fluctuations caused by the long-range tails of Coulomb
potentials of the remote ionized dopants.\cite{davi91,Topi00}

The presence of potential fluctuations is responsible for the
suppression of conductance quantization in modulation-doped wires
longer than 0.5~$\mu$m, even though the electron mean free path
is longer than 10~$\mu$m.\cite{Timp91} This has stimulated the
development of new methods, such as carrier accumulation by means
of external electrodes rather than by modulation
doping.\cite{hare99, kane98, tkac01, yaco97} Although this method
allows to observe conductance quantization in wires as long as 20
$\mu$m, however, at millikelvin temperatures they usually are
obscured by irregular and reproducible fluctuations indicating
quantum interference of electron waves scattered by residual
disorder. On the other hand, the enormous sensitivity of
conductance quantization to even minute disorder can be used for
the reciprocal purpose, namely, as a probe of nanoscale potential
fluctuations. We exploit this here to demonstrate that the unique
properties of PbTe\cite{khok03} lead to an almost total
suppression of the potential fluctuations, even in the presence
of significant concentration of charged impurities and
dislocations. We draw this conclusion from the observation of
accurately quantized conductance steps in narrow constrictions
lithographically patterned of PbTe/PbEuTe quantum wells,
containing much more defects than AlGaAs/GaAs heterostructures.
We assign this behavior to the paraelectric properties of PbTe,
which result in a Curie-like temperature dependent static
dielectric constant, approaching a huge value of $\epsilon =
1350$ at 4.2 K.\cite{baue83} Because the Coulomb potentials of
charged defects are suppressed, the conducting electrons are only
scattered by short-range potentials of defect and impurity cores,
which makes the observation of ballistic transport phenomena
possible. Thus, successful nanostructurization of this system
would bring ideal quantum devices, free of the effects of
nanoscale potential fluctuations.

Due to the lack of a perfect lattice-matched substrate and the
resulting imperfections in PbTe epitaxial growth there are some
additional sources of disorder in the system. While in our
previous works we demonstrated the presence of conductance
quantization in nanoconstrictions of wide, 50 nm thick PbTe
quantum wells,\cite{grab99, grab02, grab04} the perpendicular
quantization energy was smaller than 1 meV and thus comparable to
the lateral one. For this reason, one dimensional (1D) energy
levels were densely distributed and therefore even small
potential fluctuations cause their overlap. Additionally, due to
the oval cross-section of these constrictions, the higher 1D
states were orbitally degenerate. Thus, although conductance
steps were observed, they were neither flat nor their magnitudes
corresponded exactly to the quantized values. Nevertheless, the
small value of quantization energy together with the large
magnitude of the Land\'e factor $|g^*| \approx 66$ made it
possible to fabricate an efficient spin filter in which the spin
polarized current was transported by several waveguide modes,
even in moderate magnetic fields of the order of only
1~T.\cite{grab02} In the present work, we focus on submicron
constrictions fabricated from much narrower PbTe quantum wells,
where the two transverse quantization energies are significantly
different. In these structures we demonstrate precise conductance
quantization in $2e^2/h$ units despite significant charged
defects in the vicinity of the constrictions. Thus, these
nanostructures not only display a similar behaviour as high
quality GaAs/AlGaAs wires, but also offer access to the region of
lifted spin degeneracy of electron states already at low magnetic
field.

\section{Multilayer fabrication and properties}
The multilayers used for fabrication of the constrictions were
grown by molecular beam epitaxy (MBE) onto (111) BaF$_2$
substrate by using the protocols described in detail in
\cite{spri03}. As shown schematically in Fig.~1, in the
structures a 12~nm PbTe quantum well resides between
Pb$_{0.92}$Eu$_{0.08}$Te barriers. For this particular Eu
composition,  the barrier is as high as 235~meV. Due to the (111)
growth direction, the fourfold $L$-valley degeneracy of the
conduction band in PbTe is lifted, so that the ground-state 2D
subband is formed of a single valley with the long axis parallel
to the [111] growth direction.\cite{Yuan94} According to envelope
function calculations,\cite{Yuan94, abram01} the first excited
subband is formed of the same valley and resides at about 24~meV
above the ground state. The lowest state formed by the three
remaining valleys, obliquely oriented to the growth directions,
has a still higher energy of 35~meV above the ground state.

\begin{figure}

\includegraphics[width=9cm]{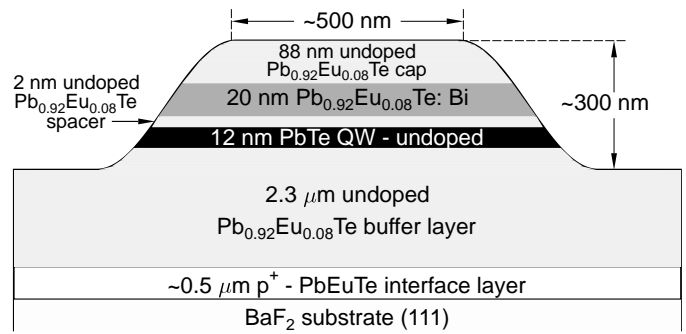}
\caption{Schematic view of the PbTe mesa cross-section showing
the layout of the initial layer grown by MBE.} \label{fig:fig1}
\end{figure}

\begin{figure}

\includegraphics[height=7.5 cm]{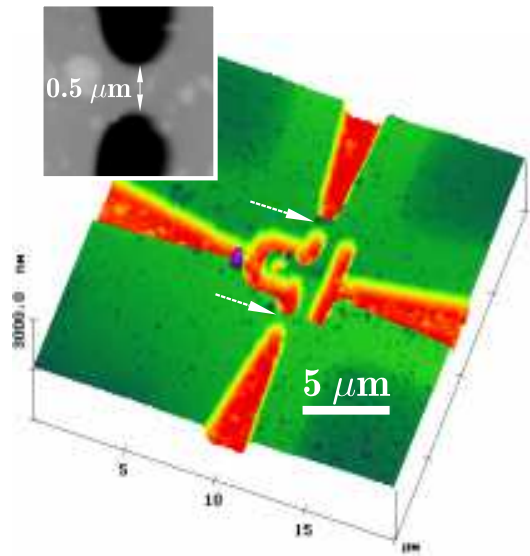}
\caption{[color on-line]  Atomic force microscopy image of a PbTe
nanostructure. The studied constrictions are marked by arrows.
Inset shows magnification of the  constriction profile.}
\label{fig:fig2}
\end{figure}

In order to introduce electrons into the quantum well, modulation
doping by Bi ($N_{\mbox{D}} \approx 3\cdot 10^{18}$~cm$^{-3}$ )
was employed with an undoped 2~nm wide Pb$_{0.92}$Eu$_{0.08}$Te
spacer layer separating the quantum well and the doping layer.
This is much thinner than that usually employed in the case of
the GaAs/AlGaAs system. Standard transport measurements reveal
total electron density $n_{\mbox{2D}} = 5.9\cdot
10^{12}$~cm$^{-2}$ and mobility $8.8\cdot 10^4$~cm$^2$/Vs in the
PbTe quantum wells at $T=4.2$~K. The analysis of Shubnikov-de
Haas oscillations shows that at least five 2D subbands are
occupied. The corresponding Fermi energy is evaluated to be at
$80 \pm 10$~meV above the bottom of ground-state subband.

In the quantum well, the electron mobility is strongly reduced
with respect to record values for the bulk-like PbTe epilayers of
up to  $2\cdot 10^6$~cm$^2$/Vs,\cite{spri03} despite the
application of the modulation doping. This results mainly from
the strong alloy scattering\cite{Prin99} at the
PbTe/Pb$_{0.92}$Eu$_{0.08}$Te interfaces. Additionally, the
system contains a significant number of threading dislocations
formed during the initial growth on the BaF$_2$ surface due to
4.2\% lattice mismatch. These dislocations act as acceptors and
account for the presence of a thin p$^+$ interfacial layer. The
application of a thick buffer layer reduces the dislocation
density which, however, remains still significant in the quantum
well region, at least at the level of $10^7$/cm$^2$.\cite{spri96}
Furthermore, the difference in thermal expansion coefficients
between the whole layer structure and the BaF$_2$ substrate
produces thermal strains of the order of 0.16\% when the
structure is cooled down to cryogenic temperatures.\cite{spri03}
There are experimental evidences that this strain induces
movement of dislocations and produces additional
defects.\cite{olve94, zogg94} Obviously, even a single of the
enlisted mechanisms would preclude the observation of conductance
quantization in a standard material. However, as we show below,
this is not the case for PbTe due to its paraelectric nature.

\section{Fabrication and properties of constrictions}

PbTe nanostructures  were fabricated by electron-beam lithography
in the form of deeply etched mesas, employing the procedures
described in our previous works.\cite{grab99,grab04} A number of
nanostructures of different forms was patterned. In the present
work, we consider only two-probe conductance of constrictions
like those marked by arrows in Fig.~2. Their width in the
narrowest region is about 0.5~$\mu$m and the total length about
1~$\mu$m. The insulating trenches have a depth of 0.3~$\mu$m.
According to our previously established procedure,\cite{grab04}
an efficient tuning of the electron concentration in the
constriction is possible by biasing the p-n junction that is
formed between the p$^+$ interfacial layer and the n-type quantum
well.

We have tested conductance of ten constrictions at 4.2~K. Six of
them were conducting and tunable by the gate voltage. Two of this
set have shown good conductance quantization, while for the
others the quantization steps were significantly distorted. A
possible origin is the presence of threading dislocations in the
quantum well region. Dislocation cores act not only as acceptors
but produce also long-range strain fields vanishing as
~$1/r$.\cite{bacon84} Although the long-range Coulomb potential
is expected to be suppressed by the lattice polarizability, the
strain field remains largely unaltered and causes electron
scattering. Therefore, the good structures are presumably those
where the active region does not contain any dislocation.
According to our previous work, the mean distance between
dislocations is around 3~$\mu$m in our samples.\cite{spri96}

We have also found that electron motion in such structures is
ballistic over the length scales at least 1~$\mu$m. We have
observed classical ballistic effects in 1.6~$\mu$m wide Hall
crosses prepared in the same way as the constrictions. In
particular, we have found a large negative magnetoresistance dip
occurring in weak magnetic fields below
50~K.\cite{beena91,grab04} This is an indication of  ballistic
transmission of the electrons between opposite contact probes.

\section{Conductance quantization}

Measurements of conductance quantization were performed by using
the standard lock-in technique. An ac voltage of typical
frequency of 129~Hz was applied to the large contact areas and
the resulting current was measured. The voltage amplitude has
been kept low enough to maintain the linear response. In Fig.~3,
we show unprocessed experimental curves representing the device
current as a function of gate voltage $V_{\mbox{g}}$ in the
absence of a magnetic field at ~2~K. Individual curve sets were
obtained during separate measurement sessions carried out in the
course of nine months since device fabrication.  Although there
are large and non-monotonic shifts of the threshold voltages in
subsequent measurements, all curves show series of regular steps
at the same current values. For $V_{\mbox{g}}$ exceeding +0.2~V,
the gated p-n junction starts to conduct precluding meaningful
measurements. Interestingly, despite the rather high electron
concentration in the unprocessed quantum well, a complete 1D
channel depletion can be easily achieved in the narrow mesas. In
some measurements, a full depletion occurs already at
$V_{\mbox{g}}$ = 0 (Fig.~3, curve set no. 5), so that the
application of a positive voltage is necessary to activate the
channel.  However, a significant depletion is observed only for
constriction widths $w$ smaller than 0.5~$\mu$m. For wider
constrictions,  the conductance remains always  high, for
instance at $w=0.65$~$\mu$m,  the conductance can be reduced only
by about 20\% even  at $V_{\mbox{g}} = - 0.9$~V, which is close
to the gate breakdown limit. In certain cases (Fig.~3 curve sets
no.~3 and 4) also conductance hystereses are observed when
$V_{\mbox{g}}$ is swept up and down. The set of results indicate
that the number and positions of defects placed close to the
constriction changes during thermal cycling and is also affected
by the junction bias.

\begin{figure}
\includegraphics[width=9.5cm]{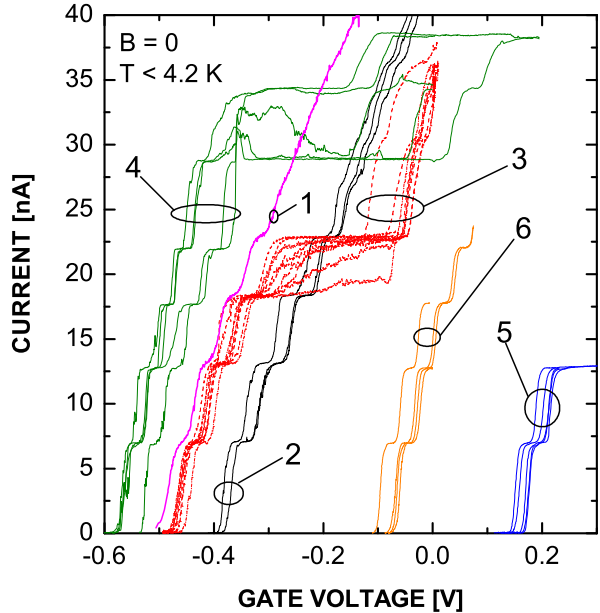}
\caption{[color on-line] Experimental traces of  the device
current as a function of the gate voltage at zero magnetic field
and at pumping helium temperatures. Particular sets of curves
indicated by numbers have been obtained during subsequent
measurement sessions.} \label{fig:fig3}
\end{figure}

Conductance traces in $2e^2/h$ units obtained in various external
magnetic fields $B$ perpendicular to the surface during session
no.~2 are summarized in Fig.~4. A series contact resistance of
140~$\Omega$ has been subtracted from the raw data. In the $B =0$
case, the four lowest conductance steps are equal to $i(2e^2/h);
(i=1,..4)$  with an accuracy better than 1\%. Although further
steps are less pronounced, one can easily resolve quantized
conductance at values corresponding to  $i = 5, 6, 8$, and 10. It
is well known that backscattering is suppressed by the magnetic
field,\cite{beena91} which should lead to an improvement of the
step flatness. In the inset to Fig.~4, the step $i = 1$ at
various fields is shown on an enlarged scale. There is no visible
improvement in the flatness up to 0.8~T (arrow) but in higher
fields the step width starts to increase.  At the same time a
gradual decrease of the step height is observed, which arises
from a contribution of the Hall effect in the macroscopic contact
pads to the total device resistance. In higher magnetic fields,
half-quantized steps start to develop, indicating the removal of
spin degeneracy.

\begin{figure}
\includegraphics[width=9.5cm]{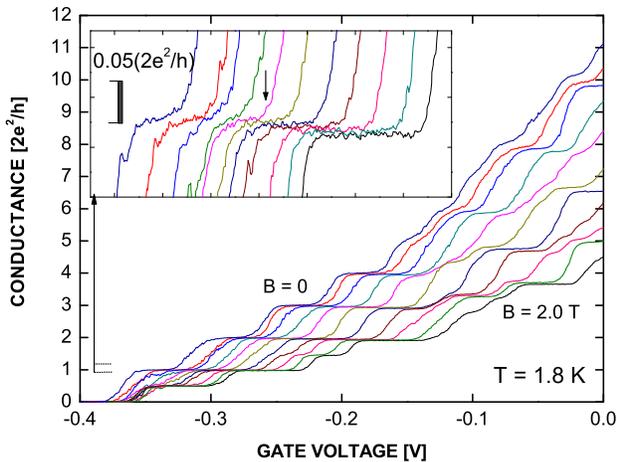}
\caption{[color on-line]Constriction conductance as a function of
$V_{\mbox{g}}$ at 1.8~K in various magnetic fields from 0 to 2~T
with a step 0.2~T. Inset shows magnification of the magnetic
field evolution of the conductance step with the index  $i=1$.}
\label{fig:fig4}
\end{figure}

It is well known that scattering by random potentials leads to
quantum interference,\cite{alts86, lee86} which shows up as
reproducible aperiodic conductance fluctuations that perturb the
quantized step structure below 1~K.\cite{vWee91, wrob92} Such
low-temperature aperiodic fluctuations are even visible  for
constrictions of ultra pure GaAs/AlGaAs heterostructures with
mobilities as high as $10^7$~cm$^2$/Vs.\cite{kane98} For this
reason, we have examined our point contacts at millikelvin
temperatures in order to probe of the effect of potential
fluctuations upon the device characteristics. The measurements
were carried out by using a $^3$He/$^4$He dilution refrigerator,
just before the measurement session no.~5 (see Fig.~3).  As shown
in Fig.~5, at 140 mK the conductance steps are even much sharper
and flatter than those observed at 1.8~K. The inset shows
magnification of the first conductance step, which reveals
reproducible conductance fluctuations but with an amplitude of
only a few percent of the total conductance.

\begin{figure}
\includegraphics[width=9.5cm]{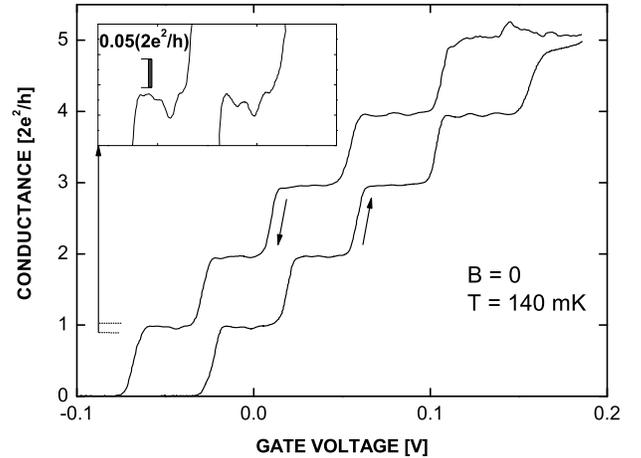}
\caption{Zero-field conductance as a function of  $V_{\mbox{g}}$
measured  at millikelvin temperatures. Arrows denote directions
of the voltage sweep. Inset shows magnification of one of the
steps. } \label{fig:fig5}
\end{figure}

\section{Discussion}

The above findings clearly indicate that the confining potential
of the electrons into 1D electron channels in PbTe is smooth at
the scale of the constriction length of about 1~$\mu$m. This
seems to contradict the observation of the irreproducible shifts
of the current traces shown in Fig.~3 that are obviously caused
by changes in the defect distributions in the system. In order to
resolve this problem, we have performed a simulation of the shape
of the 1D confining potential of the constriction geometry shown
in the inset to Fig.~2, taking into account the high dielectric
constant of PbTe. In particular, we have calculated the potential
at the narrowest part of the constriction, responsible for the
conductance quantization. We have considered the near-depletion
regime, where the the contribution of the conducting electrons to
the total potential can be neglected. This is justified because
the background doping density, reaching
10$^{17}$~cm$^{-3}$,\cite{spri03} is relatively high and just
comparable to the estimated 1D electron density in the range
where the lowest conductance steps are observed.\cite{scha79}
Then the 1D confining potential is entirely produced by
positively charged donors randomly distributed in the doped
layer, as it is illustrated schematically  in Fig.~6(a).

\begin{figure}
\includegraphics[width=6cm]{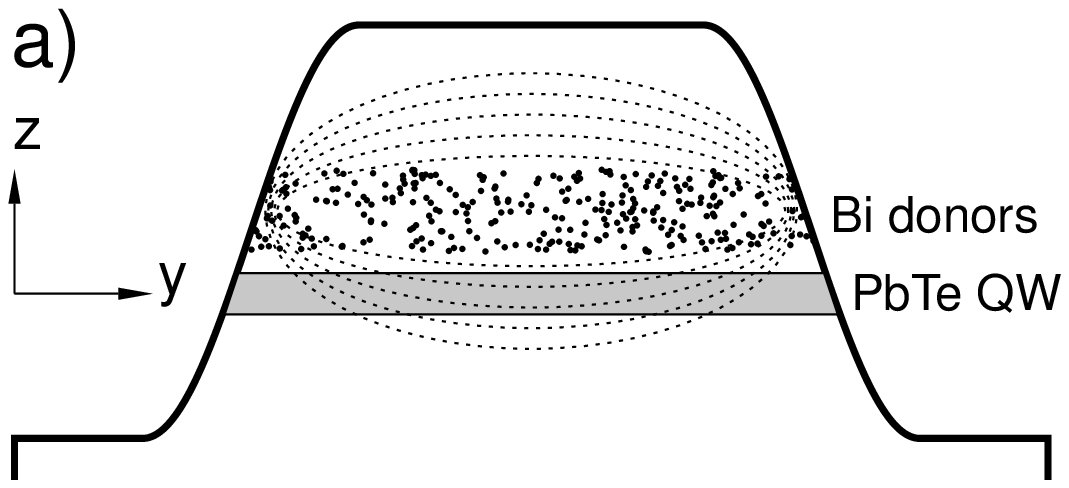}
\includegraphics[width=6cm]{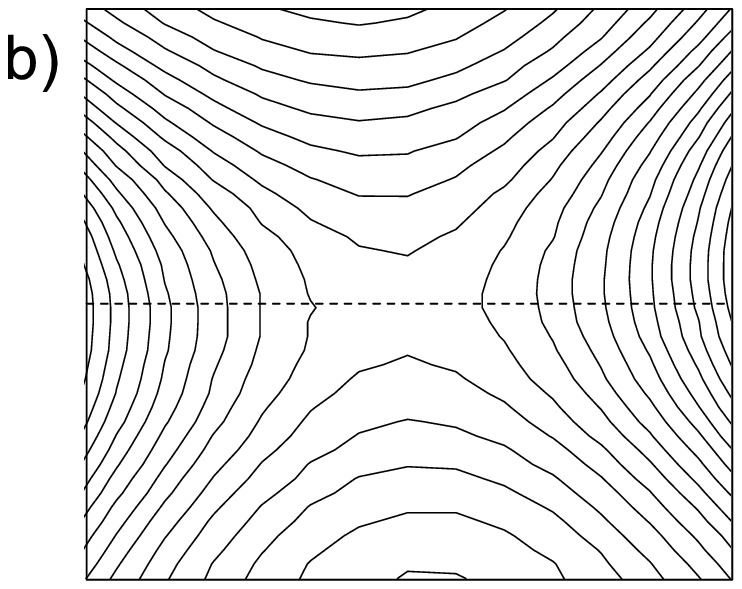}
\includegraphics[width=6cm]{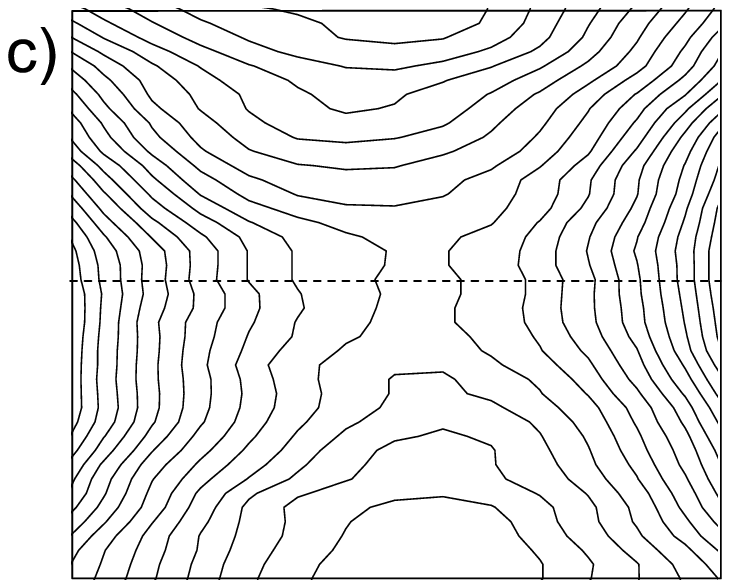}
\caption{a) Schematic picture of the constriction cross-section
used for calculation of the 1D potential profile produced by
randomly distributed donors (marked as black dots). Dotted lines
indicate equipotential contours. b) Calculated potential profile
on the square of area 0.4$\times $0.4 $\mu $m, in the x-y plane
produced by uniformly distributed donor charge in the doping
layer. Contours correspond to an energy spacing of 1~meV.  c) The
same profile calculated for the charge of randomly distributed
discrete donors. } \label{fig:fig6}
\end{figure}

Assuming a donor density of 3$\cdot$ 10$^{18}$~cm$^{-3}$, we have
calculated the potential map in the PbTe quantum well by
numerical summation of their Coulomb potentials. As a reference
point, we have calculated the potential map for continuously
distributed donor charge shown in Fig.~6(b). It represents a well
defined saddle-point potential. The calculation performed for
discrete donors also gives a saddle-point but there are visible
distortions of the potential contours (Fig.~6(c)). However, these
deviations do not exceed 1~meV. In particular, the transverse
confining potential at the narrowest cross-section (dashed lines)
is practically the same in the both cases. We find that it is
nearly parabolic with curvature corresponding to the 1D energy
level spacing $\hbar\Omega = 2.5$~meV. Importantly, this value
depends only on the width of the donor stripe located just above
the saddle-point. We have checked this by carrying out the same
calculations but with additional distributions of charges. For
example, we have assumed various randomly distributed surface
charges, volume charge corresponding to the background doping or
the non-depleted electron charge in the wider regions of the
constriction. In all these cases the potential curvature, and
thus the 1D level spacing, are practically the same. Obviously,
this remains valid if the additional charge density is smaller
than the doping donor density or its distance from the channel is
appropriately large. The only effect produced by the additional
charge is a rigid shift of the whole 1D energy level structure.

Our simulations appears to explain pertinent experimental
features. In particular, the calculated magnitude of the
potential fluctuations is comfortably smaller than the 1D level
spacing. In fact, the presence of screening by conducting
electrons in real channels will reduce the potential fluctuations
even more. Furthermore, the persistence of conductance
quantization despite disorder results from the robustness of the
parabolic potential curvature to the presence of randomly
distributed charged defects. Because their potentials contribute
merely to a constant potential background, they cause only a
shift of the threshold voltage without deteriorating the step
structure. In particular, the observed initial depletion of the
channel is presumably caused by a large concentration of
negatively charged defects existing in the constriction vicinity
and/or on the device surfaces. One of possible candidates are
negatively charged acceptors generated by dislocations. The
significant shifts of the threshold voltage between different
cooling sessions indicate a dominant role of the thermal stress
in the defect redistribution, in accordance with previous
magnetotransport measurements of 2D PbTe quantum well
structures.\cite{olve94} Furthermore, the presence of hystereses
when $V_{\mbox{g}}$  is swept up and down indicates that the
defect distribution is affected by the gate electric field that
is of the order of $10^3$--$10^4$~V/cm.

As shown in Fig.~4, a near perfect conductance quantization is
observed only for the steps with quantum numbers smaller than 5.
This is because these 1D levels have an energy below the first
excited 2D subband in the initial quantum well. For $i>4$,
intersubband mixing starts to appear and we return to the
situation encountered in the oval PbTe constrictions, where 1D
energy levels exhibit orbital degeneracies.\cite{grab04} From our
calculations, the fifth 1D quantum level is at an energy,
$4.5\hbar\Omega = 11$~meV, a value about two times smaller in
comparison with calculated energy of the first excited subband.
However, it should be recalled that for the potential simulations
we have taken the geometrical profile (Fig.~1 and 2) for the
width of the donor layer. If, for any possible reasons this layer
is narrower, $\hbar\Omega$ will increase. For example, one could
suspect the existence of some donor compensation at the wire
edges.

Our investigations in magnetic fields confirm the lack of
potential fluctuations because there is no quantization
improvement due to suppression of backscattering by the field.
However, because of the small in-plane effective electron mass
$m^*=0.02m_0$, the Landau splitting becomes larger than
$\hbar\Omega$ already at $B=0.5$~T. Then the magnetic
quantization dominates over the 1D wire quantization, and this
causes a substantial widening of the steps. Similarly to other 1D
systems, the steps gradually evolve into quantum Hall effect
plateaux\cite{beena91} with increasing widths.

Finally, it worth mentioning that although the long range tails
of the Coulomb potentials are strongly suppressed, the
short-range cores of the scattering potentials remain effective.
Since the MBE growth of PbTe results in background donor
concentration of the order of 10$^{16}$--10$^{17}$~cm$^{-3}$,
there are about 10 to 100 short-range scattering centers within
the channel. Our data shows, however, that they do not destroy
the conductance  quantization. One can recall here a number of
theoretical models describing the conductance of 1D constrictions
containing short-range scattering centers, approximated by Dirac
$\delta$-function.\cite{bagw90, tekm90, boes00, varg05} The
common result of these models is the strong dependence of the
conductance quantization steps on the sign of the scattering
potentials. In particular, for repulsive centers, the steps are
altered rather weakly. However, the presence of attractive
scatterers is predicted to produce large resonance dips
superimposed on the steps. This remains valid even when extending
the theory to scattering centers of small, but non-zero,
range.\cite{varg05} Because we do not have any independent
information on the sign of the short-range part of impurity
potentials in PbTe, the problem of their possible residual
influence on conductance quantization remains open.

\section{Summary}

In conclusion, we have observed {\sl precise} zero-field
conductance quantization in submicron constrictions of PbTe
quantum wells embedded in Pb$_{1-x}$Eu$_x$Te barriers, similar to
GaAs/AlGaAs quantum wires. We find a robustness of the
conductance quantization against charged defects in the
constriction vicinity, in a stark contrast to any other known
systems. This results from the stressing of nanoscale Coulomb
potential fluctuations by the huge static dielectric constant of
PbTe. As a consequence, charge defects do not scatter the
carriers but only shift the threshold voltage. At the same time,
we do not see conductance resonances expected for scattering by
short-range potentials, which may indicate that either the
corresponding scattering cross sections are too small or that
their potential is repulsive. Our results demonstrate the
suitability of PbTe nanostructures as promising system for
quantum devices.

\section{Acknowledgments}

This work was partly supported by PBZ/KBN/044/P03/2001 Grant,
ERATO Semiconductor Spintronics Project, as well as the
Gesellschaft fuer Mikroelektronik and the Fonds zur Foerderung
der wissenschafftlichen Forschung of Austria.

\end{document}